\documentclass{mn2e}
\usepackage{epsfig}

\title[K~{\sc i}~$\lambda$7699 in the nebula of VY CMa]{Spatially
extended K~{\sc i}~$\lambda$7699 emission in the nebula of VY~CMa:
Kinematics and geometry}

\author[N.\ Smith]{Nathan Smith\thanks{Hubble Fellow;
nathans@casa.colorado.edu}\thanks{Visiting Astronomer at the New
Technology Telescope of the European Southern Observatory, La Silla,
Chile.} \\ Center for Astrophysics and Space Astronomy, University of
Colorado, 389 UCB, Boulder, CO 80309, USA}

\date{Accepted 0000, Received 0000, in original form 0000}
\pagerange{\pageref{firstpage}--\pageref{lastpage}} \pubyear{2002}

\begin{document}
\label{firstpage}
\maketitle
\begin{abstract}

Long-slit echelle spectra reveal bright extended emission from the
K~{\sc i} $\lambda$7699 resonance line in the reflection nebula
surrounding the extreme red supergiant VY Canis Majoris.  The central
star has long been known for its unusually-bright K~{\sc i} emission
lines, but this is the first report of intrinsic emission from K~{\sc
i} in the nebula.  The extended emission is not just a reflected
spectrum of the star, but is due to resonant scattering by K atoms in
the outer nebula itself, and is therefore a valuable probe of the
kinematics and geometry of VY~CMa's circumstellar environment.
Dramatic velocity structure is seen in the long-slit spectra, and most
lines of sight through the nebula intersect multiple distinct velocity
components.  A faint ``halo'' at large distances from the star does
appear to show a reflected spectrum, however, and suggests a systemic
velocity of $+$40 km s$^{-1}$ with respect to the Sun.  The most
striking feature is blueshifted emission from the filled interior of a
large shell seen in images; the kinematic structure is reminiscent of
a Hubble flow, and provides strong evidence for asymmetric and
episodic mass loss due to localized eruptions on the star's surface.

\end{abstract}

\begin{keywords}
circumstellar matter --- stars: evolution --- stars: individual (VY
CMa) --- stars: mass-loss
\end{keywords}

\section{INTRODUCTION}

Due to enhanced mass loss during the final stages of stellar
evolution, red supergiants are surrounded by extended gas and dust
envelopes.  This material forms the pre-supernova circumstellar
environment, or may get swept-up to form a ring nebula if the star
evolves blueward to become a luminous blue variable or Wolf-Rayet
star.  In a brief phase when mass-loss peaks, these stars can have
circumstellar nebulae dense enough to emit OH masers and strong
thermal infrared (IR) emission from dust that may obscure the star
(OH/IR stars).  Among the most luminous red supergiants above
10$^{5.5}$ L$_{\odot}$, only a handful in our Galaxy are OH/IR stars
--- one of the most luminous is the M5e~Ia supergiant VY Canis
Majoris.  At a distance of 1.5 kpc (Herbig 1972; Lada \& Reid 1978;
Marvel 1997), its luminosity is 10$^{5.7}$ L$_{\odot}$.

VY CMa is a special case among the most luminous red supergiants,
because of its spectacular reflection nebula, with strong polarization
(Herbig 1972) and intricate knotty and filamentary structure seen in
{\it Hubble Space Telescope} ({\it HST}) and ground-based IR images
(Kastner \& Weintraub 1998; Monnier et al.\ 1999; Smith et al.\ 2001).
These studies revealed a chaotic and highly asymmetric distribution of
dust in the circumstellar environment.  VY CMa also has strong SiO,
H$_2$O, and OH maser emission (Knowles et al.\ 1969; Buhl et al.\
1974).  Proper motions and Doppler shifts of these masers have
repeatedly been interpreted to suggest that VY CMa is surrounded by an
expanding disk-like distribution of material with the polar axis
oriented northeast-southwest (van Blerkom \& Auer 1976; Rosen et al.\
1978; Benson \& Mutel 1979, 1982; Morris \& Bowers 1980; Bowers et
al.\ 1983; Deguchi et al.\ 1983; Richards et al.\ 1998).  These maser
studies only probe material close to the star, and seem hard to
reconcile with the severely asymmetric arrangement of knots, arcs, and
chaotic filaments seen at larger radii in optical {\it HST} images
(Smith et al.\ 2001; see Figure 1), or the asymmetric structure near
the star in the IR (Monnier et al.\ 1999).  A disk-like geometry
implies axial symmetry, so perhaps the receding part of the nebula is
obscured by foreground material.  Nebulae around red supergiants are
neutral dusty reflection nebulae, without convenient diagnostics of
the geometry and kinematics like the bright H$\alpha$ or [N~{\sc ii}]
lines commonly used in studies of nebulae around hot stars.

%% Figure 1
\begin{figure*}\begin{center}
\epsfig{file=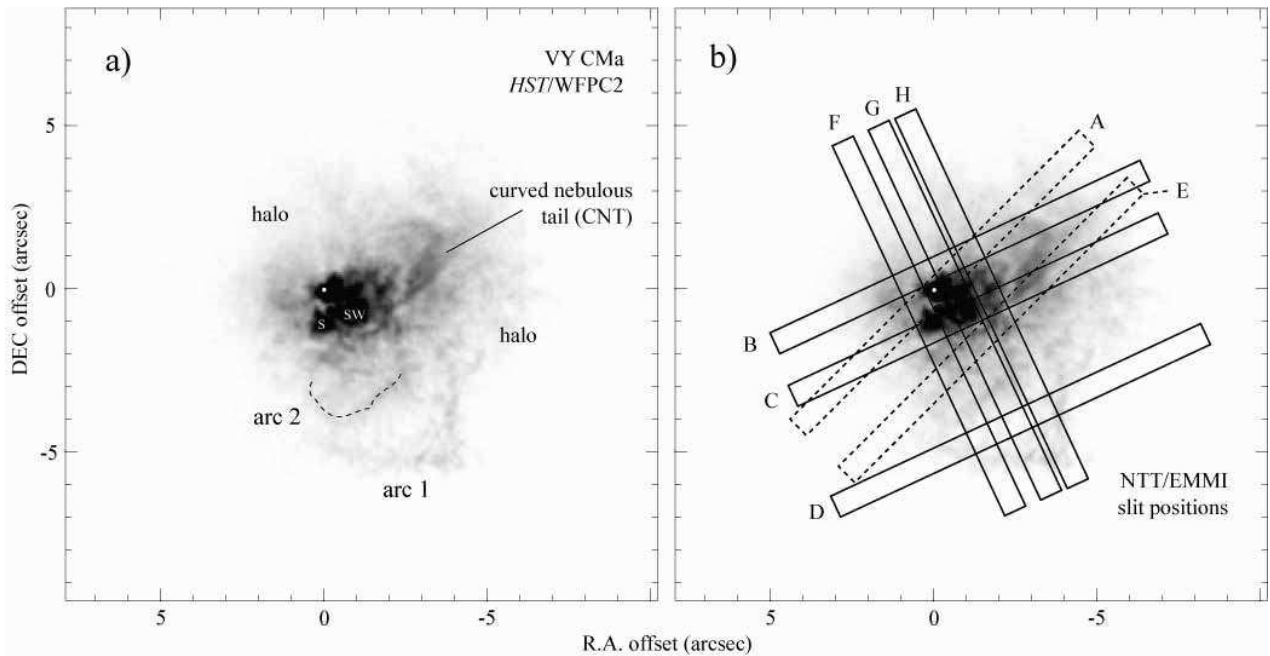,width=6.6in}\end{center}
\caption{ Visual-wavelength {\it HST}/WFPC2 image of VY CMa from Smith
et al.\ (2001), except that here unsharp masking has been applied.
Several features discussed in the text are labeled in (a), and (b)
shows the positions and orientations of EMMI long-slit aperture
positions superposed on the same image of the nebula.  Letters A
through H correspond to position-velocity diagrams in individual
panels of Figure 2, where the position of the letter relative to the
slit corresponds to the top of each panel.  The small white dot marks
the position of the star at visual wavelengths, but the true position
of the star is offset slightly to the northeast (Smith et al.\ 2001;
Kastner \& Weintraub 1998).  Color images showing the outer stucture
in more detail were published by Smith et al.\ (2001).}
\end{figure*}

Spectroscopically, VY CMa is unique among red supergiants because the
resonance lines of K~{\sc i} $\lambda$7665 and $\lambda$7699 are seen
strongly in emission (Wallerstein 1958; Humphreys 1970; Wallerstein \&
Gonzalez 2001), accompanied by P Cygni absorption.  The K~{\sc i}
lines arise in cool gas at $\sim$700 to 1000 K (Wallerstein 1958),
comparable to dust temperatures near the star (Monnier et al.\ 1999;
Smith et al.\ 2001).  Its spectrum is also unusual in that it shows
emission from TiO, ScO, Ti~{\sc i}, Cr~{\sc i}, Rb~{\sc i}, Ba~{\sc
ii}, etc.\ (Joy 1942; Wallerstein 1958, 1986; Hyland et al.\ 1969;
Wallerstein \& Gonzalez 2001).  Emission from the K~{\sc i} lines is
sometimes seen in extended envelopes of red supergiants, including
Betelgeuse (Bernat \& Lambert 1976; Bernat et al.\ 1978; Mauron et
al.\ 1984; Plez \& Lambert 1994, 2002), but the extreme strength of
K~{\sc i} emission in the spectrum of the central star is rare.

This Letter reports the discovery that the previously-known K~{\sc i}
emission in the star's spectrum is accompanied by bright {\it
extended} K~{\sc i} emission from VY~CMa's nebula.  The K~{\sc i}
lines are by far the brightest emission lines throughout the IR and
visual-wavelength spectrum of the nebula.  K~{\sc i} emission traces
roughly the same spatial extent as the reflection nebula, and yields
interesting clues to the kinematics and geometry of VY CMa's
circumstellar environment.

%% Table 1
\begin{table}
\caption{Spectroscopic Observations of VY CMa}
\begin{tabular}{@{}lcccl}	\hline\hline

Slit &Date &P.A. &Exp. &Comment \\ &d/m/y &(deg) &(min) & \\ \hline

A	&8/3/03		&135	&9	&star		\\
B	&11/3/03	&295	&30	&offset N, CNT	\\
C	&11/3/03	&295	&30	&offset S, CNT	\\
D	&11/3/03	&295	&30	&offset S, arc1	\\
E	&8/3/03		&135	&30	&CNT and arc2	\\
F	&10/3/03	&205	&30	&offset E, arc2	\\
G	&10/3/03	&205	&9	&offset W, knots, arc1 \\
H	&10/3/03	&205	&30	&offset W, arc1	\\ \hline

\end{tabular}
\end{table}

%% Figure 2
\begin{figure*}\begin{center}
\epsfig{file=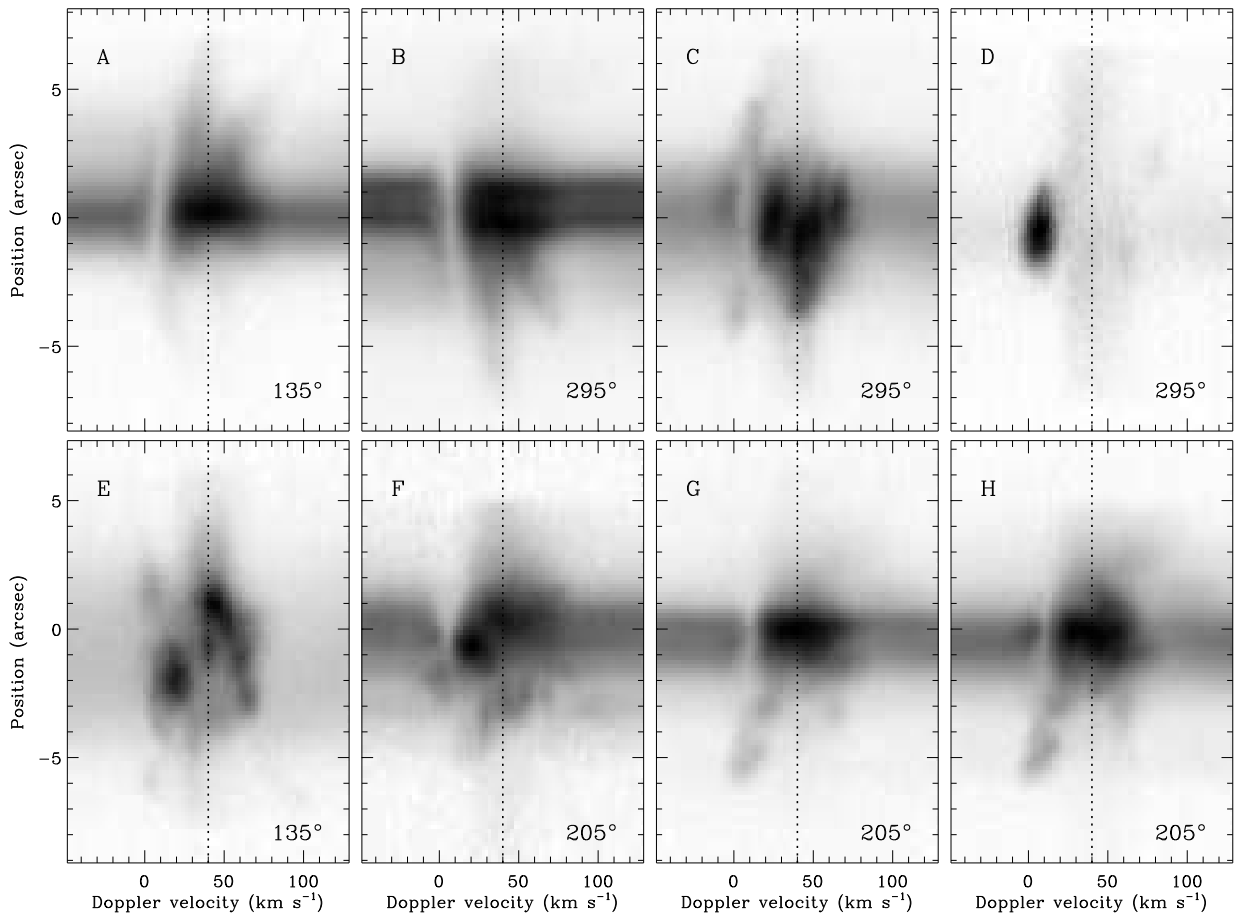,width=6.7in}\end{center}
\caption{
Position-velocity diagrams of K~{\sc i}~$\lambda$7699 emission from
the nebula of VY CMa.  Panels A through H correspond to slit positions
A through H labeled in Figure 1$b$.  The upward direction on each
panel in this figure corresponds to the position of the label A
through H in Figure 1$b$, and down in each panel corresponds to the
position angle given in the lower right and listed in Table 1.
Heliocentric Doppler velocity is indicated on the X-axis, and the
dashed vertical line marks the presumed systemic velocity of $+$40 km
s$^{-1}$.  The full slit length of $\sim$13$\arcsec$ is slightly
shorter than the range of position shown here.}
\end{figure*}

\section{OBSERVATIONS}

High-resolution optical spectra of VY CMa were obtained at the
European Southern Observatory (ESO) at La Silla, Chile on the nights
of 8, 10, and 11 March 2003 using the ESO Multi-Mode Instrument (EMMI)
on the New Technology Telescope (NTT).  Cross-dispersed echelle
spectra were obtained using EMMI's red CCD (a 4096$\times$4096 pixel
MIT/LL detector) with the number 10 echelle grating and grism 6 for
order separation.  The separation between adjacent orders on the CCD
allows for some spatial information to be recorded with a
$\sim$13$\arcsec$ long slit aperture; this is well-suited to VY~CMa's
compact reflection nebula (Figure 1).  On all three observing nights
the seeing was quite good, varying from 0$\farcs$6 to 0$\farcs$8 on
average.  To sample this good seeing and to maximize spectral
resolution, a narrow 0$\farcs$8-wide slit aperture was used.  This
configuration yielded a spectral resolution of $R = \lambda /
\Delta\lambda \; \approx \; 35,000$ (8 km s$^{-1}$), and a pixel scale
of 0$\farcs$166 $\times$ 0.0517 \AA \ (or $\sim$2 km s$^{-1}$ at a
wavelength of 7699 \AA).  Wavelength calibration was accomplished
using observations of an internal emission lamp; heliocentric
velocities for the K~{\sc i} $\lambda$7699 line are quoted here.
Several different slit positions and orientations were used to sample
the spectrum of various features seen in images, as shown in Figure
1$b$.  Discussion in this Letter is limited to the velocity structure
of the very bright K~{\sc i} $\lambda$7699 emission line, since the
other K~{\sc i} line at 7665 \AA \ is strongly affected by absorption
in the atmospheric A-band, and no other line in the spectrum had
comparable brightness in the ejecta.  Figure 2 shows the resulting
position-velocity diagrams for K~{\sc i} $\lambda$7699 at the various
slit orientations in Figure 1$b$.

%% Figure 3
\begin{figure}\begin{center}
\epsfig{file=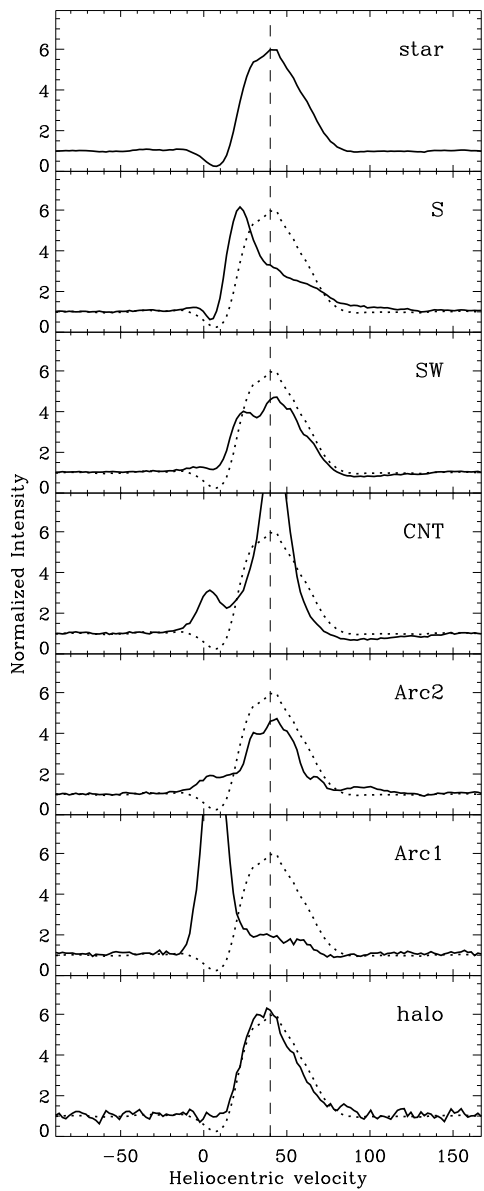,width=2.65in}\end{center}
\caption{ Intensity tracings of K~{\sc i} $\lambda$7699 at selected
positions in the nebula.  Names correspond to features labeled in
Figure 1$a$; the ``halo'' spectrum is the average of several different
positions in the nebula sampled from near the ends of the slit.  All
spectra have been normalized, for comparison with the star's spectrum
(shown with a dotted line in each panel).  The heliocentric systemic
velocity of $+$40.0 km s$^{-1}$ is shown with a dashed line.}
\end{figure}

\section{DISCUSSION}

The observations presented here were exploratory, and the most
significant result is the discovery of bright {\it intrinsic} K~{\sc
i} emission from the nebula around VY CMa --- this extended K~{\sc i}
emission is not simply a dusty reflection of the star's K~{\sc i}
emission line, but rather, it is intrinsic emission from the gas in
the nebula itself.\footnotemark\footnotetext{This is immediately
apparent because some of the emission is blueshifted, whereas
reflection by expanding dust can only cause redshifted emision.  A
salient example is the Homunculus nebula around $\eta$ Carinae (Smith
et al.\ 2003).}  It is therefore a powerful diagnostic of the
kinematics and geometry of the extended nebula.  Inferring quantities
like column densities and mass-loss rates from the K~{\sc i} emission
requires detailed modeling beyond the scope of this short Letter (see
the models of K~{\sc i} emission for the nebula of Betelgeuse; Jura \&
Morris 1981; Rodgers \& Glassgold 1991; Glassgold \& Huggins 1986),
but the information provided here can hopefully be useful in such an
investigation.  A qualitative description of the data is given here,
and some clues to the kinematics and geometry of the nebula are
discussed.  Figure 3 shows spectral intensity tracings at a few
selected positions, and Table 2 collects some measurements from these
tracings, such as the flux-weighted centroid velocity ($v_{\rm cen}$),
the emission equivalent width (EW), the flux ratio of reflected
continuum to the central star's continuum ($F$/$F_*$), and the
intensity ratio of the extended K~{\sc i} emission compared to the
emission line seen in the star (I/I$_*$).  Uncertainty in $I$/$I_*$,
$F$/$F_*$, and the EW is a few percent, or as much as $\pm$10\% in the
faintest outer ``halo'' spectrum.  The values for EW, $F$/$F_*$, and
$I$/$I_*$ in Table 2 were corrected for instrumental scattering using
observed values of $F$/$F_*$ for a point source.

%% Table 2
\begin{table}
\caption{Measurements from spectral tracings}
\begin{tabular}{@{}lccccc}	\hline\hline

Feature	&Radius		&$v_{\rm cen}$	&EW	&F/F$_*$ &I/I$_*$	\\
	&(arcsec)	&(km s$^{-1}$)	&(\AA)	&(cont.) &(line)	\\ \hline

star	&...		&+41.5		&4.75	&1	&1		\\
S	&1		&+32.4		&4.02	&0.13	&0.11		\\
SW	&1		&+41.0		&3.33	&0.10	&0.07		\\
CNT	&3		&+39.2		&6.53	&0.024	&0.033		\\
Arc2	&3		&+40.1		&3.96	&0.030	&0.025		\\
Arc1	&5		&+7.7		&5.27	&0.0083	&0.0092		\\
halo	&6		&+40.0		&4.66	&0.0014	&0.0014		\\ \hline

\end{tabular}
\end{table}

\subsection{Kinematic Structure}

Examining Figures 2 and 3, it is obvious that the kinematic structure
of K~{\sc i} $\lambda$7699 in VY~CMa's circumstellar nebula is
complex, and not easily explained by spherical or axial symmetry.
This is consistent with qualitative inspection of {\it HST} images of
the nebula, which suggest either highly asymmetric mass ejection or
selective extinction (or both; Smith et al.\ 2001; Kastner \&
Weintraub 1998).  In general, bright K~{\sc i} emission with strong
Doppler shifts is confined to within about 5$\arcsec$ from the star,
and strong emission features can usually be associated with knots or
filaments in images of the nebula (Figure 1).  Many positions show
multiple velocity components along the same line of sight, especially
the bright knots like S and SW.  The curved nebulous tail (CNT; see
Herbig 1972) shows interesting kinematic structure (see slit position
E in Figure 2, which runs along the CNT).  Near the star it is
redshifted by about 20 km s$^{-1}$ (compared to $v_{\rm sys}$), and
this Doppler shift gradually decreases to match $v_{\rm sys}$ at the
apex of the CNT farthest from the star.  Thus, the CNT might either be
a bubble like Arc1 moving near the plane of the sky, or an expanding
equatorial ring.  Overall, the material with strong K~{\sc i} emission
and chaotic kinematic structure is seen mainly toward the southwest of
the star, in the brightest part of the reflection nebula.

Self absorption is not uniformly distributed either.  Blueshifted
absorption is seen in the bright central parts of the nebula near the
star, and is concentrated toward the north and northeast.  Ratio
images also showed enhanced reddening toward the northeast (Smith et
al.\ 2001).  No absorption is seen more than 1$\arcsec$ from the star
toward the southwest direction.  The narrow absorption shows a
velocity gradient with position: It shifts from about $+$20 km
s$^{-1}$ (heliocentric) at a few arcseconds northwest of the star,
reaching almost 0 km s$^{-1}$ at positions a few arcsec east of the
star.

While the faint outer halo has an emission profile almost identical to
that of the star (Figure 3), the P Cyg absorption component is
missing.  The K~{\sc i} equivalent width in the halo is also nearly
identical to the star (Table 2), so at large radii this halo is
probably a pure reflection nebula.  However, the missing P Cyg
absorption component indicates that other lines of sight do not see
the P Cyg absorption.  Either the stellar wind of VY~CMa is
asymmetric, or the absorption is not a ``real'' P Cyg component formed
in the wind, but is instead localized self absorption in the nebula.

\subsection{Systemic velocity}

At more than 5$\arcsec$ from the star in most position-velocity
diagrams in Figure 2, a very faint emission component is seen at
nearly the same velocity as that of the star.  Images of VY~CMa reveal
a faint, fairly uniform scattering halo extending beyond 5$\arcsec$
from the star as well (Smith et al.\ 2001).  Figure 3 shows an average
of several different positions in this outer halo.  This is not simply
instrumental scattering of the bright star, because the P Cyg
absorption is missing in the outer parts.  This outer halo may be the
best tracer of the systemic velocity of VY~CMa.  Table 2 gives the
flux-weighted centroid velocity of K~{\sc i} in the halo, indicating a
likely systemic velocity of $+$40.0$\pm$1.5 km s$^{-1}$
(heliocentric), or $+$21 km s$^{-1}$ (LSR).  This agrees with various
other indicators of VY~CMa's systemic velocity, varying between $+$37
and $+$44 km s$^{-1}$ (Wallerstein 1986; Bowers et al.\ 1983; Deguchi
et al.\ 1983; Reid \& Dickinson 1976; Neufeld et al.\ 1999; Harwit \&
Bergin 2002), and is close to the systemic velocity of molecular
clouds associated with the nearby cluster NGC~2362 ($+$18 km s$^{-1}$
LSR; Lada \& Reid 1978).

\subsection{Episodic ejection}

Deep {\it HST} images of VY~CMa revealed a filamentary arc about
5$\arcsec$ southwest of the star (Arc 1 in Figure 1$a$) that provoked
speculation about asymmetric, episodic mass ejection due to stellar
activity (Smith et al.\ 2001).  Slit positions D, G, and H in Figure
1$b$ were placed to study the velocity structure of Arc 1, and the
results are striking.  K~{\sc i} emission associated with Arc 1 is
strongly blueshifted.  Interestingly, it seems to show a quasi-Hubble
law, perhaps composed of several clumps seen best in Figure 2 G and
H. Its blueshift increases with separation from the star, reaching a
maximum of about $-$40 km s$^{-1}$ with respect to the star at a
separation of 5$\arcsec$. (The emission feature at $-$3$\arcsec$ and
$+$60 km s$^{-1}$ in Figure 2 G and H may be Arc 2, consistent with
the high reddening of Arc 2 in images; see Smith et al.\ 2001.)  A
Hubble-like flow would imply a single ejection episode, strengthening
the hypothesis that Arc 1 is due to an asymmetric eruption from the
star's surface.

The pseudo `Hubble constant' for this flow is roughly $-$7.2 km
s$^{-1}$ arcsec$^{-1}$. Without proper motions, the projection angle
out of the plane of the sky $\psi$ is unknown, but at a distance of
1.5 kpc, the observed velocity structure implies an age of $t \approx
1000 / (\tan \psi)$ years.

The spatial extent of the K~{\sc i} emission associated with Arc 1
suggests that the emission line originates mainly in the filled {\it
interior} of a bubble outlined by Arc 1.  Perhaps reflected light in
images traces a thin filamentary dust shell swept up by the expanding
gas emitting K~{\sc i}.

\subsection{Comparison with Betelgeuse}

Neutral potassium emission has been observed in the circumstellar
environment of Betelgeuse as well, out to radii of roughly 50$\arcsec$
(Plez \& Lambert 2002), whereas the brightest K~{\sc i} reaches
roughly 5$\arcsec$ from VY~CMa.  Since VY~CMa is about 10 times
farther away, the actual radial extent of the K~{\sc i} emission is
comparable around these two stars.  However, the K~{\sc i} in VY~CMa's
circumstellar shell is much brighter with respect to the star; I/I$_*$
is a factor of 10$^3$ to 10$^4$ stronger in VY~CMa.  Note that
Betelgeuse shows no K~{\sc i} emission in the direct stellar spectrum,
so Plez \& Lambert measured $I_*$ as the continuum flux in a 1-\AA\
bin; thus, $I$/$I_*$ in Table 2 needs to be multiplied by a factor of
4.75 (the EW of K~{\sc i} $\lambda$7699 for the star) for direct
comparison with Betelgeuse.  This relatively bright extended K~{\sc i}
emission is partly due to the fact that direct photospheric light from
VY~CMa is obscured by its own circumstellar dust (Smith et al.\ 2001;
Kastner \& Weintraub 1998; Herbig 1972).  Perhaps even the ``stellar''
K~{\sc i} emission arises in unresolved circumstellar ejecta, but
appears so strong because the star is occulted.  This may help explain
the mystery of the unusually-bright K~{\sc i} emission from VY~CMa.
However, understanding the physical conditions in the gas that gives
rise to the bright K~{\sc i} emission, the potassium abundance, the
mass-loss rate, and other properties will require quantitative
modeling of the observed nebular spectrum.

\smallskip\smallskip\smallskip\smallskip
\noindent {\bf ACKNOWLEDGMENTS}
\smallskip
\scriptsize 

\noindent Support was provided by NASA through grant HF-01166.01A from
the Space Telescope Science Institute, which is operated by the
Association of Universities for Research in Astronomy, Inc., under
NASA contract NAS 5-26555.  I thank an anonymous referee for helpful
comments.

\label{lastpage}
\end{document}